\documentclass[11pt,twoside]{article}
\usepackage{asp2004}
\usepackage{psfig}
\usepackage{epsf}
\usepackage{graphics}
\usepackage{lscape}

\reversemarginpar

\begin{document}
\newcommand{\msun}{\mbox{M$_{\odot}$}}
\title{The Most Massive Stars in the Local Group: Measuring Accurate
Masses of Stars in Eclipsing Binaries}
\author{Alceste Z. Bonanos, Krzysztof Z. Stanek}
\affil{Harvard-Smithsonian Center for Astrophysics \\ 60 Garden Street,
Cambridge, MA 02138}

\begin{abstract}

Accurate masses and, in general, all fundamental parameters of distant
stars can only be measured in eclipsing binaries. Several massive star
candidates with masses near $200\msun$ exist, however they have large
uncertainties associated with them. The most massive binary ever
measured accurately is WR~20a, for which we present the light
curve. Measuring the period and inclination, we derive masses greater
than $80\msun$ for each component. Massive binaries are bound to exist
in Local Group galaxies, such as M31 and M33. These can be selected
from their light curves, obtained by variability studies, such as the
DIRECT project. We present photometry and spectroscopy of the detached
system M33A, for which we are obtaining a direct distance
determination. The DIRECT project has detected several candidate
massive binaries which are brighter but non-detached systems, perhaps
similar to WR~20a. We plan to obtain spectra for them and measure
their masses.

\end{abstract}
\thispagestyle{plain}

\section*{Introduction}

Measuring accurate masses for the most massive stars in our Galaxy and
beyond is important for constraining star formation and stellar
evolution theories, which have indirect implications for many objects
that are not well understood, such as supernovae, gamma-ray bursts and
Population III stars. Some of the most massive candidates in the Milky
Way are LBV 1806-20 \citep{Eikenberry04,Najarro04}, the Pistol Star
\citep{Figer98}, and $\eta$ Carinae \citep{Davidson97}, which have
inferred masses up to $\sim 200\;$\msun. However, the masses of these
stars are only indirect estimates and thus have large uncertainties
associated with them. The only direct way of measuring accurate masses
of distant stars is in eclipsing binaries.

Until recently, the most massive stars ever weighed in binaries were:
R136-38 (O3V+O6V) in the LMC with a primary mass of $56.9\pm 0.6
\;\msun$ \citep{Massey02}, WR 22 (WN7+abs$\;+\;$O), with a minimum
primary mass of $55.3\pm 7.3\; \msun$ \citep{Rauw96,Schweickhardt99}
and Plaskett's star with a minimum primary mass of $51\; \msun$
\citep{Bagnuolo92}. We present the light curve for WR~20a, the most
massive binary measured accurately, and propose to measure fundamental
parameters of similar massive binaries in the Local Group.

\section*{The New Heavyweight Champion WR~20a}

The current champion of the most massive star competition is WR~20a, a
Wolf-Rayet (WR) binary in the compact cluster Westerlund~2, which is
located at the center of the HII region RCW 49. \citet{Rauw04}
obtained spectroscopy for WR~20a and measured extremely large minimum
masses of $70.7 \pm 4.0$ and $68.8 \pm 3.8\;\msun$ for the
components. The final masses strongly depend on both the period and
inclination of the binary, which can only be measured from the light
curve. In May 2004, \citet{Bonanos04} obtained 83 $I-$band
observations of WR~20a with the OGLE team's 1.3 m Warsaw telescope at
Las Campanas Observatory, Chile, which is operated by the Carnegie
Institute of Washington.

\begin{figure}[!ht]  
\plotfiddle{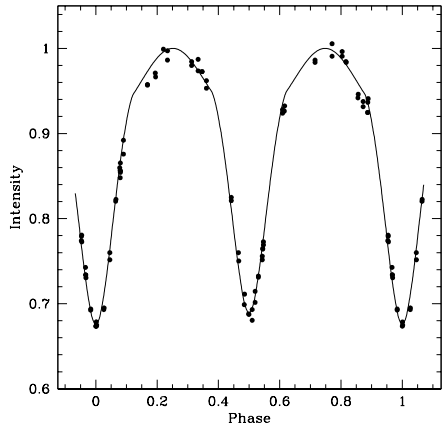}{5cm}{0.0}{30}{30}{-100}{-50}
\caption{Wilson-Devinney model fit of a near contact binary to the
$I$-band light curve of WR~20a. The period is $3.686$ days, the
eccentricity is $0$ and the inclination angle $i=74.5 \deg$.}
\label{lc}
\end{figure} 

The first goal of our observations was to confirm the $\sim4$ day
spectroscopic period of the WR~20a binary and refine its value to
$3.686$ days, thus confirming the remarkable masses of its
components. Next, we derived an inclination angle $i$ for the system
from our well-sampled light curve. We fit the curve with the
Wilson-Devinney (WD) code \citep{Wilson71,Wilson79,vanHamme03} for
modeling distorted stars and derived a best fit inclination angle of
$i=74.5\pm 2.0\deg$. In Figure~\ref{lc} we show the result of the
$I-$band light curve model fit for WR~20a. The uneven eclipse depths
suggest slightly different effective temperatures and, thus, different
spectral types for the components.

Armed with the refined period and the exact value of the inclination
angle, we re-analyzed the radial velocity data of \citet{Rauw04}.
Fixing the eccentricity to $0$, as the light curve confirms, and the
period to $3.686$ days, we fit the radial velocity data applying equal
weights and rejecting points with radial velocity measurements smaller
than $80\;\rm km\;s^{-1}$. We derive slightly larger velocity
semi-amplitudes than \citet{Rauw04}, which in turn yield larger
minimum masses and final masses of the components of $83.0 \pm 5.0\;
\msun$ and $82.0 \pm 5.0\; \msun$.

\section*{Massive Binaries in the Local Group}

\begin{figure}[!ht]  
\plottwo{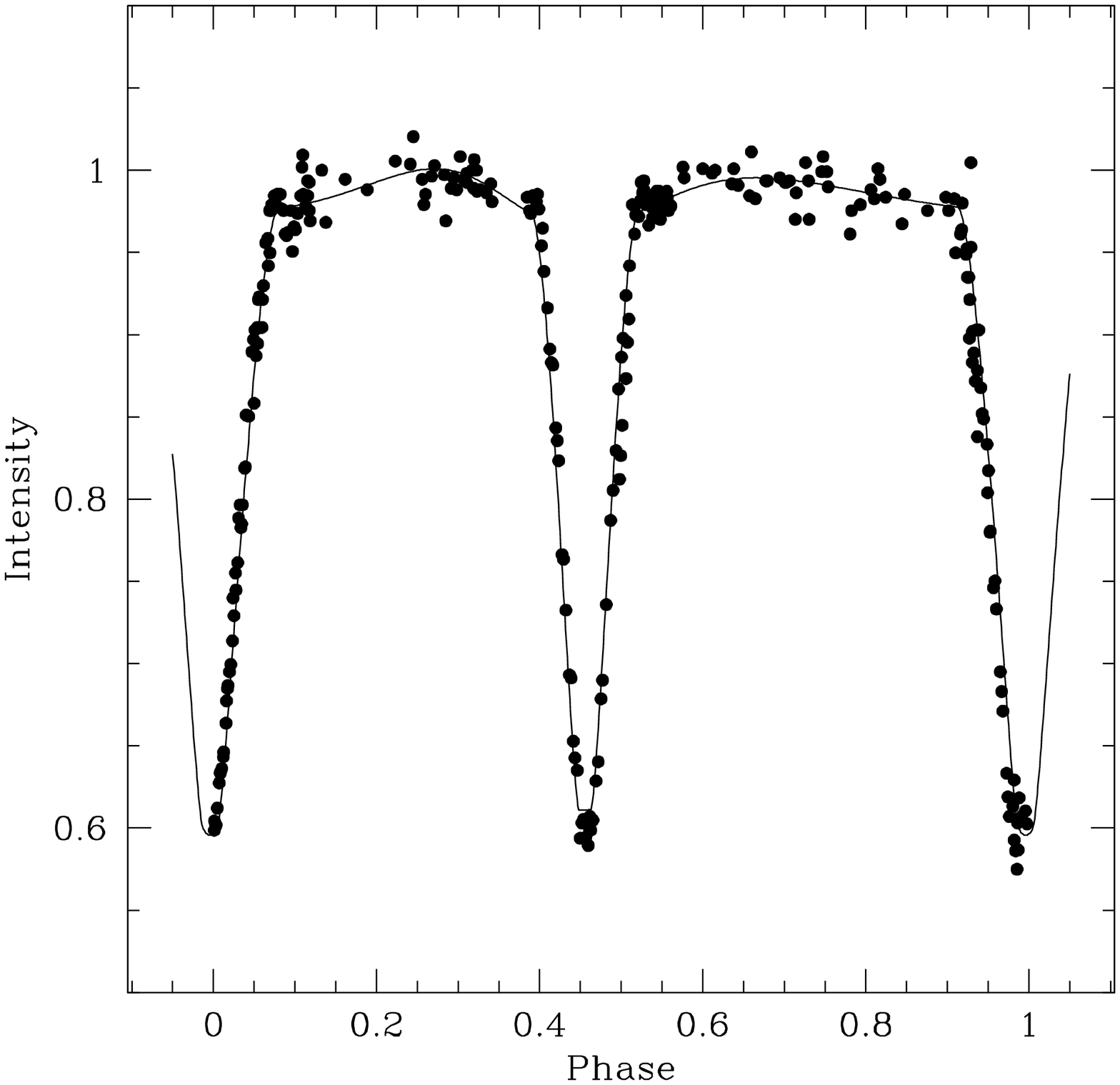}{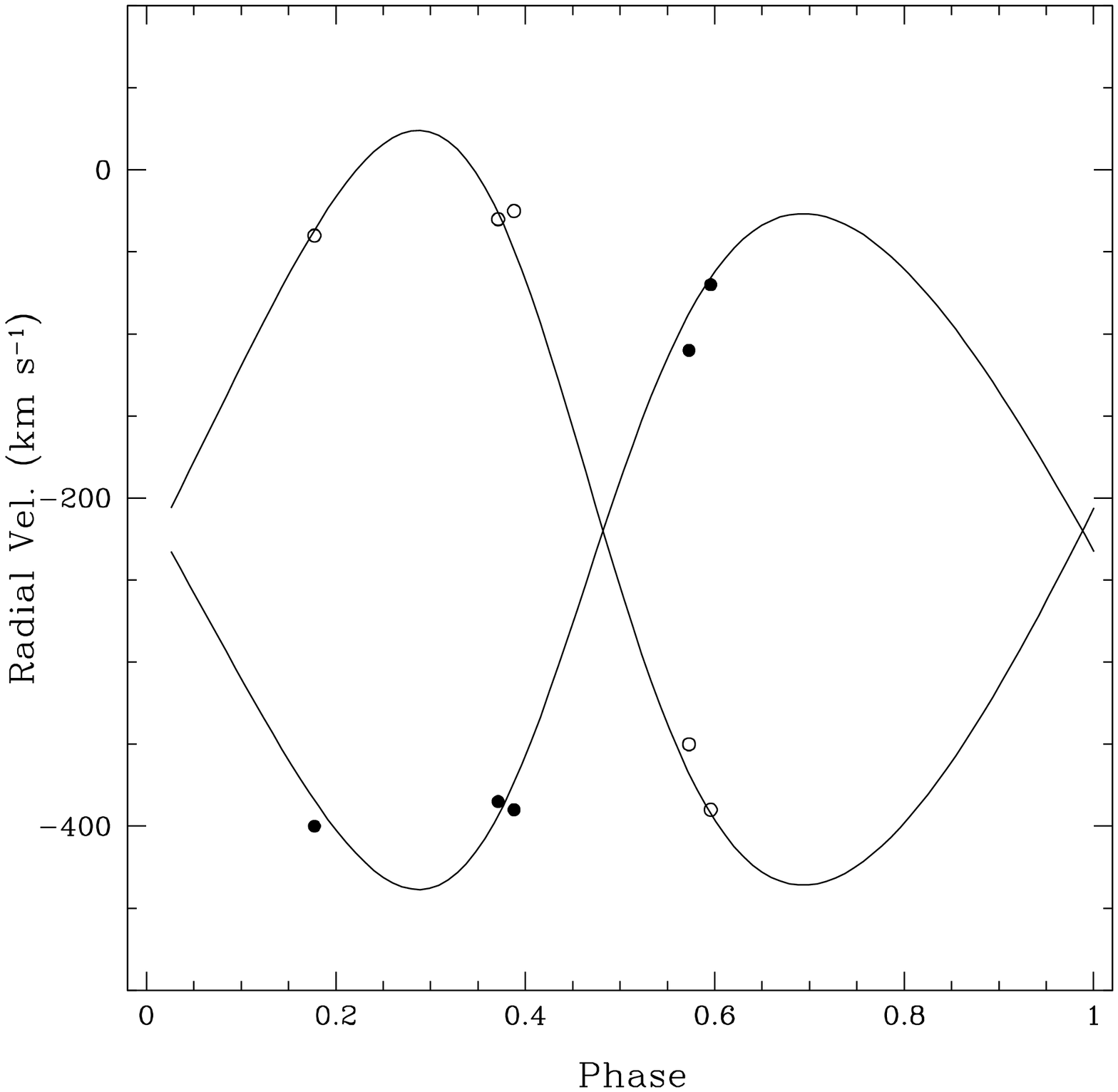}
\caption{Phased light curve and radial velocity curve of the detached
binary M33A. Photometry was obtained by the 2.1 m at KPNO and
spectroscopy with ESI on Keck II.}
\label{DEB}
\end{figure}

The DIRECT Project \citep[e.g.][]{Stanek98, Bonanos03} has found
$\sim1000$ variables in each of the galaxies M31 and M33, which
include about 100 eclipsing binary systems. Follow up photometry was
done with the 2.1m KPNO telescope in 1999 and 2001 on the best system,
M33A, and the light curve is shown on the left panel of
Figure~\ref{DEB} (A.Z. Bonanos et al. 2004, in preparation). We have
obtained spectra with ESI on Keck II in November 2002 and September
2003 and the radial velocity curve is shown on the right panel of
Figure~\ref{DEB}.

Analogs of WR~20a, possibly with even higher mass components, are
bound to exist in M31 and M33. We have selected the brightest
eclipsing binaries as massive binary candidates and plan to obtain
spectroscopy to derive their radial velocity curves and, hence,
masses. In addition to measuring their fundamental parameters, we will
obtain a direct distance estimate. The advantage of studying eclipsing
binaries in these galaxies would be their well known distances and low
reddening.

\vspace{0.3cm} 
We would like to acknowledge our collaborators Guillermo Torres, Rolf
Kudritzki, Andrzej Udalski and Dimitar Sasselov for contributing to
these projects.


\newpage
\begin{thebibliography}

\bibitem[{{Bagnuolo} {et~al.}(1992){Bagnuolo}, {Gies}, \&
{Wiggs}}]{Bagnuolo92} {Bagnuolo}, W.~G., {Gies}, D.~R., \& {Wiggs},
M.~S. 1992, \apj, 385, 708

\bibitem[{{Bonanos} {et~al.}(2003){Bonanos}, {Stanek}, {Sasselov},
  {et~al.}}]{Bonanos03}
{Bonanos}, A.~Z., {Stanek}, K.~Z., {Sasselov}, D.~D., {et~al.} 2003,
  \aj, 126, 175

\bibitem[Bonanos et al.(2004)]{Bonanos04}
Bonanos, A.Z., Stanek, K.Z., Udalski, A. et al. 2004, ApJL, in press
(astro-ph/0405338)

\bibitem[{{Davidson} \& {Humphreys}(1997)}]{Davidson97}
{Davidson}, K. \& {Humphreys}, R.~M. 1997, \araa, 35, 1

\bibitem[{{Eikenberry} {et~al.}(2004){Eikenberry}, {Matthews}, {LaVine},
  {et~al.}}]{Eikenberry04}
{Eikenberry}, S.~S., {et~al.} 2004, \apj, in press (astro-ph/0404435)

\bibitem[{{Figer} {et~al.}(1998){Figer}, {Najarro}, {Morris},
  {et~al.}}]{Figer98}
{Figer}, D.~F., {Najarro}, F., {Morris}, M., {et~al.} 1998, \apj, 506,
384

\bibitem[{{Massey} {et~al.}(2002){Massey}, {Penny}, \&
{Vukovich}}]{Massey02}
{Massey}, P., {Penny}, L.~R., \& {Vukovich}, J. 2002, \apj, 565, 982

\bibitem[Najarro(2004)]{Najarro04}
Najarro, F. 2004, in \it Fate of the Most Massive Stars: title, \rm
Humphreys R. \& Stanek K.Z., eds, (San Francisco: ASP), ASP Conf. Ser
\#\#,p.\#\#

\bibitem[{{Rauw} {et~al.}(2004){Rauw}, {De Becker}, {Naze},
{et~al.}}]{Rauw04}
{Rauw}, G., {De Becker}, M., {Naze}, Y., {et~al.} 2004, \aap, 420, L9

\bibitem[{{Rauw} {et~al.}(1996){Rauw}, {Vreux}, {Gosset},
{et~al.}}]{Rauw96}
{Rauw}, G., {Vreux}, J.-M., {Gosset}, E., {et~al.} 1996, \aap, 306, 771

\bibitem[{{Schweickhardt} {et~al.}(1999){Schweickhardt}, {Schmutz}, {Stahl},
  {et~al.}}]{Schweickhardt99}
{Schweickhardt}, J., {Schmutz}, W., {Stahl}, O., {et~al.} 1999, \aap,
347, 127

\bibitem[{{Stanek} {et~al.}(1998){Stanek}, {Kaluzny}, {Krockenberger},
  {et~al.}}]{Stanek98}
{Stanek}, K.~Z., {Kaluzny}, J., {Krockenberger}, M., {et~al.} 1998, \aj, 115,
  1894

\bibitem[{{van Hamme} \& {Wilson}(2003)}]{vanHamme03}
{van Hamme}, W. \& {Wilson}, R.~E. 2003, in ASP Conf. Ser. 298: GAIA
  Spectroscopy: Science and Technology, 323

\bibitem[{{Wilson}(1979)}]{Wilson79}
{Wilson}, R.~E. 1979, \apj, 234, 1054

\bibitem[{{Wilson} \& {Devinney}(1971)}]{Wilson71}
{Wilson}, R.~E. \& {Devinney}, E.~J. 1971, \apj, 166, 605

\end{thebibliography}
\end{document}